\documentclass[preprint,preprintnumbers,amsmath,amssymb,12pt]{revtex4}

\usepackage{graphicx}
\usepackage{epsfig}
\usepackage{bm}
\usepackage{amsfonts}
\usepackage{subfigure}
\usepackage{multirow}
\usepackage{color}
\usepackage{float}
\usepackage{xcolor, soul}

\usepackage{array}
\newcolumntype{C}[1]{>{\centering\arraybackslash}p{#1}}

\usepackage{hyperref}
\hypersetup{colorlinks=True,citecolor=blue}

\graphicspath{{./figs/}}

\sethlcolor{green}

\def\thickhline{\noalign{\hrule height1.9pt}}

\begin{document}


\title{\textbf{Loop quantum inflation with inverse volume corrections in light of ACT data}}

\author{Farough Parvizi\footnote{faroughparvizi1986@gmail.com}, Soma Heydari\footnote{s.heydari@uok.ac.ir}, Milad Solbi\footnote{miladsolbi@gmail.com}, and Kayoomars Karami\footnote{kkarami@uok.ac.ir}}
\affiliation{\small{Department of Physics, University of Kurdistan, Pasdaran Street, P.O. Box 66177-15175, Sanandaj, Iran}}

\date{\today}

\begin{abstract}
Within the framework of loop quantum cosmology (LQC), we investigate the effect of inverse volume corrections on the low scale spontaneously broken supersymmetric (SB SUSY) and exponential inflationary potentials. The LQC modifications to the Friedmann equations and cosmological perturbation parameters are employed to assess the observational viability of these models against recent data from the Atacama Cosmology Telescope (ACT). Our results indicate that in contrary to the standard model of inflation, in the presence of inverse volume corrections in LQC, the prediction of SB SUSY and exponential potentials in the $r-n_{\rm s}$ plane lie inside the 68\% confidence level interval of the ACT data.
\end{abstract}


\maketitle

\section{Introduction}
Inflation is a well-recognized paradigm that resolves several shortcomings of the Hot Big Bang cosmology, like the horizon and flatness problems \cite{Guth:1981,Linde:1982}. During this accelerated expansion, quantum fluctuations of the inflaton field generate the scalar and tensor perturbations, which in turn produce the anisotropies observed in the Cosmic Microwave Background (CMB). The first step in analyzing any inflationary model is to compare its predictions for the scalar spectral index $n_{\rm s}$ and the tensor-to-scalar ratio $r$ with observational CMB data and constrain the mode parameters.

Over the past decades, measurements of CMB anisotropies have been progressively refined through a series of observational missions, from WMAP to the high-precision, full-sky measurements of the Planck satellite \cite{Planck:2018inflation,Planck:2018inflation1,bk18}.
As a result, the classical predictions of several theoretically well-motivated models, such as those arising from low-scale spontaneously broken supersymmetry (SB SUSY) \cite{Dvali:1994,Copeland:1994,Linde:1994,Ashrafzadeh:2025} and the exponential potential \cite{Lucchin:1985,Halliwell:1987,Yokoyama :1987,Burd:1988,Rezazadeh:2017,Amani:2018}, were found to be in significant tension with observations.
Nevertheless, recent data from ground-based experiments like the Atacama Cosmology Telescope (ACT) \cite{ACT:DR6models,ACT:DR6params}, combined with Planck, DESI BAO, and BICEP/Keck observations, offer a new perspective.
In this regard, the recent observational data indicates a slight upward shift in the best-fit value of the scalar spectral index, $n_s$.
While this development brings the predictions of these disfavored models tantalizingly closer to the new observational contours, most of their parameter space remains in tension.
Intriguingly, for a specific number of e-folds, such as $N=50$ in the SUSY model, the prediction now falls within the 95\% CL of the latest data.
This promising shift provides strong motivation to re-examine these specific models and explore physical mechanisms that could fully reconcile them with observations.
In response to the evolving data, the literature now features a range of theoretical approaches designed to bring inflationary models facing observational tension back into agreement with the latest constraints \cite{Kallosh:2025act,Aoki:2025,Berera:2025,Dioguardi:2025,Raidal:2025,Salvio:2025,Antoniadis:2025,Kim:2025,Drees:2025,Haque:2025,Peng:2025bws,Liu:2025qca,Gialamas:2025ofz,Wolf:2025ecy,Dioguardi:2025mpp,Heidarian:2025drk}.

In this context, it is timely to explore additional physical effects that may reconcile these models with the latest cosmological observations. Loop Quantum Cosmology (LQC), an application of Loop Quantum Gravity to cosmological settings, provides a promising theoretical framework for this purpose. LQC predicts the existence of quantum corrections to the classical equations of motion, which stem from the discrete nature of spacetime at the Planck scale.
Two commonly discussed types of effective modifications are holonomy and inverse volume corrections
\cite{Zhu:2015,Zhu:2016,Barrau:2014,Calcagni:2011,Thiemann:1998,Bojowald:2001b,Bojowald:2001,Ashtekar:2006,Bojowald:2011,Bojowald:2011c}.

Holonomy corrections emerge from the quantization procedure where the gravitational connection is represented through holonomies along finite loops.
A major consequence is the resolution of the Big Bang singularity, which is replaced by a non-singular quantum bounce occurred at a critical energy density \cite{Bojowald:2001,Ashtekar:2006}.
Although their impact is significant near Planckian scales, these corrections are generally subdominant during the slow roll phase of inflation. This leaves the background dynamics largely consistent with the standard inflation \cite{Barrau:2014}.
Inverse volume corrections, on the other hand,  arise from the quantization of operators corresponding to inverse powers of the volume. They are potentially more relevant for slow roll phase of inflation \cite{Bojowald:2001b,Thiemann:1998}.
These corrections modify the effective Friedmann and Klein-Gordon equations which lead to corrections in the evolution of cosmological perturbations \cite{Bojowald:2011,Calcagni:2011,Zhu:2015,Zhu:2016}. Moreover, they reform the dispersion relation of primordial fluctuations, which can induce scale-dependent deviations in the scalar and tensor power spectra, particularly at large scales \cite{Zhu:2015,Bojowald:2011c}. Such characteristics render a potential observational tool to probe the quantum effects predicted by LQC.

%
%

In this paper, we adopt a semi-classical formalism wherein, while the LQC corrections to the background evolution are sub-dominant during the slow roll phase, the corrections to the perturbation equations can still have a significant and observable impact, particularly on the spectral indices and their runnings. This provides a consistent framework in which to test these quantum effects against observational data.
The aim of this study is to investigate the observational viability of the SB SUSY and exponential inflationary potentials within the LQC framework with inverse volume corrections. We assess whether the modifications to the primordial power spectra predicted by LQC can improve the agreement between these models and the latest observational data from ACT. This analysis will be carried out using the most accurate analytical formulas for the LQC corrected inflationary observables available in the literature \cite{Bojowald:2011,Calcagni:2011,Zhu:2015,Zhu:2016}.

This paper is organized as follows: In Section \ref{sec2}, the basic outline of the LQC framework with inverse volume corrections is introduced. Sections \ref{sec:SUSY} and \ref{sec:exp} are devoted to analyzing the SB SUSY and exponential potentials, respectively, and confronting their LQC corrected predictions with observational data. Finally, the main conclusions of the paper are summarized in Section \ref{sec:conclusion}.

\section{Background equation and perturbations in LQC}\label{sec2}

The standard action for a canonical scalar field inflation model is given as follows \cite{Guth:1981,Linde:1982}
\begin{equation}\label{eq:action}
S=\int{\rm d}^{4}x \sqrt{-g} \left[\frac{M_{\rm p}^2}{2}R+ X - V(\varphi) \right] ,
\end{equation}
where  $M_{\rm p} \equiv 1 / {\sqrt{8\pi G}}$ is the reduced Planck mass, $R$ is the Ricci scalar, and $g$ is the determinant of the metric tensor $g_{\mu\nu}$. In addition, $V(\varphi)$ and $X\equiv\frac{1}{2}g^{\mu\nu}~\partial_\mu \varphi \partial_\nu \varphi$  denote the  potential and kinetic energy term for the inflaton field, respectively.

In LQC, inverse volume corrections modify the classical dynamics. For a flat FRW metric in conformal time given by $ ds^2=a^{2}(\tau) \left(-d\tau^2+dx^{i}dx_{i} \right)$, the LQC effective Friedmann and Klein-Gordon equations are \cite{Bojowald:2011}
\begin{align}
	&\mathcal{H}^2 = \frac{8\pi G}{3} \alpha\left[\frac{1}{2\nu}
	(\varphi')^2 + pV(\varphi) \right], \label{eq:FR1-ML}\\
	&	\varphi'' + 2\mathcal{H}\left(1-\frac{d\ln
		\nu}{d\ln p}
	\right)\varphi' +
	\nu pV_{,\varphi} = 0, \label{eq:KL-GO}
\end{align}
where primes denote derivatives with respect to conformal time  $\tau$, $p=a^2$ and $\mathcal{H}=\frac{a'}{a}$.
In these equations, the quantum corrections are encapsulated in the functions $\alpha$ and $\nu$
\begin{equation}
	\alpha \approx 1 + \alpha_0 \delta_{\rm pl},
\end{equation}
\begin{equation}
	\nu \approx 1 + \nu_0 \delta_{\rm pl},
\end{equation}
where $\delta_{\rm pl} \equiv \left(\dfrac{p_{\rm pl}}{p}\right)^{\sigma/2} =\left(\dfrac{a_{\rm pl}}{a}\right)^\sigma$ is the evolving quantum correction parameter and depends on the scale factor $a$. The parameters $\sigma$, $\alpha_0$, $\nu_0$ and $p_{\rm pl}$ are constants that depend on the specific parametrization of the loop quantization. For consistency, all quantities are expanded to the first order of $\delta_{\rm pl}$.
It should be noted that in the absence of inverse volume corrections ($\delta_{\rm pl}=0$), we have $\alpha=\nu=1$, and Eqs.~(\ref{eq:FR1-ML})-(\ref{eq:KL-GO}) return to the standard background equations of inflation.

The inflationary slow roll dynamics are characterized by the Hubble slow roll parameters, which are defined directly from the evolution of the Hubble parameter $\mathcal{H}$ and the scalar field $\varphi$ as follows
\begin{align}
\epsilon &\equiv 1 - \frac{\mathcal{H}'}{\mathcal{H}^2},\label{eq:SRH1}\\
\eta &\equiv 1 - \frac{\varphi''}{\mathcal{H}\varphi'}.\label{eq:SRH2} 
\end{align}
Slow roll inflation occurs when these parameters are small, \emph{i.e.}, $(|\epsilon|, |\eta|) \ll 1$.
These dynamical parameters can be related to the geometry of the inflaton potential, $V(\varphi)$, through the standard potential slow roll parameters
$\epsilon_{\rm V}\equiv \frac{M_p^2}{2} \left( \frac{V'}{V} \right)^2$ and $\eta_{\rm V}\equiv M_p^2 \left(\frac{V''}{V}\right)$. In the presence of LQC inverse volume corrections, the relationship between these two sets of parameters is modified. To the first order in the quantum correction term $\delta_{\rm pl}$, the Hubble parameters are approximated by \cite{Calcagni:2011}
\begin{align}
\epsilon &\approx\epsilon_{\rm V}+ \left\{\frac{\sigma\alpha_0}{2}-\left[\alpha_0\left(1-\sigma\right)
+\nu_0\left(\frac{\sigma}{2}-1\right)\right]\epsilon_{\rm V}
-\frac{\sigma\alpha_0}{3}\eta_{\rm V} \right\} \delta_{\rm pl},\label{eq:SRH1}\\
\eta &\approx \eta_{\rm V}-\epsilon_{\rm V} \nonumber \\
& \quad -\left\{ 
\sigma\left(\frac{\alpha_0}{2}+\frac{\sigma \nu_0}{3}-\nu_0 \right)
+\left[\alpha_0\left(\sigma-1\right)+\nu_0\left(1-\frac{7\sigma}{6}+\frac{\sigma^2}{9}\right)\right]\epsilon_{\rm V}\right. \nonumber \\ 
& \quad \left. 
+\left[\alpha_0\left(1-\frac{\sigma}{2}\right)+\nu_0\left(\frac{2\sigma}{3}-1\right)\right]\eta_{\rm V} \right\}\delta_{\rm pl}.\label{eq:SRH2} 
\end{align}
In the presence of inverse volume corrections in LQC, the power spectra of scalar and tensor perturbations are modified. Under the slow roll approximation, one can show that the scalar and tensor power spectrum at horizon crossing $k=\mathcal{H}$ take the following forms \cite{Calcagni:2011}

\begin{eqnarray}
	&\mathcal{P}_s\simeq\dfrac{\mathcal{H}^2}{8 \pi ^{2}M_{\rm p}^{2} \epsilon} (1 + \gamma_s \delta_{pl})\Big|_{k=\mathcal{H}},\label{eq:Ps-SR} \\
	&\mathcal{P}_t\simeq\dfrac{2\mathcal{H}^2}{\pi ^{2}M_{\rm p}^2} (1 + \gamma_t \delta_{pl})\Big|_{k=\mathcal{H}} \label{eq:Pt-SR}.
\end{eqnarray}
The correction coefficients $\gamma_s$ and $\gamma_t$ are given by
\begin{equation}
	\gamma_s =
	\nu_0 \left(\frac{\sigma}{6} + 1
	\right) + \frac{\sigma\alpha_0}{2\epsilon} - \frac{\chi}{\sigma+1},
\end{equation}
\begin{equation}
	\gamma_t = \frac{\sigma-1}{\sigma+1} \alpha_0,
\end{equation}
with
\begin{equation}
	\chi \equiv \frac{\sigma
		\nu_0}{3} \left(\frac{\sigma}{6} + 1
	\right) + \frac{\alpha_0}{2}\left(5-\frac{\sigma}{3}
	\right).
\end{equation}
Planck measurements constrain the scalar power spectrum to ${\cal P}_{s}(k_\ast) \simeq 2.1 \times 10^{-9}$ at the pivot scale $k_\ast = 0.05~{\rm Mpc^{-1}}$ \cite{Planck:2018inflation,bk18}.
Using these expressions, the scalar spectral index $n_{\rm s}$
and  the tensor-to-scalar ratio $r$ are found to be
\begin{equation}
		n_{\rm s} - 1 \equiv \frac{d\ln \mathcal{P}_{\rm s}}{d\ln k}\approx - 6\epsilon_V + 2\eta_V - c_{\rm n_{\rm s}} \delta_{pl}, \label{eq:ns_in_LQC}
\end{equation}
\begin{equation}
	r\equiv \dfrac{\mathcal{P}_t}{\mathcal{P}_s}\approx 16\epsilon_V + c_{\rm r} \delta_{pl}, \label{eq:r_in_LQC}
\end{equation}
where the coefficients $c_{n_{\rm s}}$ and $c_{\rm r}$ are functions of the standard potential slow roll parameters and LQC parameters as follows
\begin{equation}\label{eq:cns_in_LQC}
	c_{\rm n_{\rm s}} = f_{\rm s} - \left[6\alpha_0 (1-\sigma) -
	\nu_0 \left(6 - \frac{13\sigma}{3} + \frac{2\sigma^2}{9}
	\right)
	\right]\epsilon_V - \left[\alpha_0 \left(\frac{7\sigma}{3} - 2
	\right) + 2
	\nu_0 \left(1 - \frac{2\sigma}{3}
	\right)
	\right]\eta_V,
\end{equation}
\begin{equation} \label{eq:cr_in_LQC}
	c_{\rm r} = \frac{8\left[3\alpha_0 (3+5\sigma+6\sigma^2 ) -
		\nu_0 \sigma(6+11\sigma)
		\right]}{9(\sigma+1)} \epsilon_V - \frac{16\sigma\alpha_0}{3} \eta_V,
\end{equation}
with
\begin{equation}
	f_{\rm s} = \frac{\sigma\left[3\alpha_0 (13\sigma-3) +
		\nu_0 \sigma(6+11\sigma)
		\right]}{18(\sigma+1)}. \label{eq:fs_in_LQC}
\end{equation}
The newest observational constraint on the scalar spectral index has been established by the combined data from ACT DR6, Planck 2018, DESI BAO, and BICEP/Keck as  $n_s = 0.974 \pm 0.003$~\cite{ACT:DR6params,ACT:DR6models}. Also,  the most recent data from Planck and BICEP/
Keck 2018 imposes an upper limit on $r$ of $r < 0.036$ \cite{bk18}.

\subsection{LQC inverse volume parameters and model analysis}

The LQC Inverse-Volume model is described by the parameters $\alpha_{0}$, $\nu_0$, and $\sigma$,
which are subject to quantization ambiguities. In order to create a more predictive framework, the number of free parameters can be reduced by employing a consistency relation derived from the requirement of an anomaly-free constraint algebra. For $\sigma\neq 3$, this relation links $\nu_0$ and $\alpha_0$ as follows
\begin{equation} \label{eq:alfa0_nu0_in_LQC}
	\nu_0 = \dfrac{3 (\sigma-6)}{(\sigma+6)(\sigma-3)} \alpha_0.
\end{equation}
This allows the effect of the inverse volume corrections to be characterized by two primary quantities: the exponent $\sigma$ and a single composite parameter $\delta(k_0)$ that represents the amplitude of the quantum correction at a given pivot scale $k_0$. For $\sigma\neq 3$, this parameter is defined as
\begin{equation} \label{eq:delta(k_0)1_in_LQC}
	\delta(k_0) =\alpha_0 \delta_{pl}(k_0).
\end{equation}
In the special case where $\sigma=3$, the consistency condition requires $\alpha_0=0$, and the effective parameter is defined in terms of $\nu_0$ instead
\begin{equation} \label{eq:delta(k_0)2_in_LQC}
\delta(k_0) =\nu_0 \delta_{pl}(k_0).
\end{equation}
It has been shown that smaller values of $\sigma$ are theoretically preferable and for $\sigma \geq 2$ quantum gravitational effects become undetectable \cite{Bojowald:2011,Calcagni:2011}.
In order to conduct a comprehensive study of the quantum gravitational effects on our model, we perform our analysis over the parameter range $\sigma \in [0,3).$
In the following sections, we will utilize this LQC framework to investigate the SB SUSY and exponential potentials. We adopt a semi-classical approach where the inverse volume corrections to the background inflationary dynamics are considered negligible, as their effect is sub-dominant during the slow roll phase. However, even when small, these quantum corrections can still introduce significant and potentially observable effects at the level of the cosmological perturbations, particularly on the spectral indices. The primary objective is to analyze how inverse volume corrections, parameterized by $\sigma$ and $\delta$, influence the key inflationary observables, such as $n_{\rm s}$ and $r$. This analysis will be carried out in light of recent observational data, particularly from the ACT DR6 release, to assess whether incorporating quantum gravity corrections improves the consistency of these models with high-precision measurements.

\section{SB SUSY potential in LQC}\label{sec:SUSY}

We first analyze the spontaneously broken supersymmetric (SB SUSY) potential originated from particle physics, given by \cite{Planck:2018inflation1,Dvali:1994,Copeland:1994,Linde:1994}
\begin{equation}\label{eq:susy_pot}
  V(\varphi) = V_0 \left[1+\alpha \ln\left(\frac{\varphi}{M_{\rm p}}\right)\right],
\end{equation}
where $ V_0$, with the dimension of $M_{\rm p}^4$, can be obtained from fixing the scalar power spectrum at the pivot scale ${\cal P}_{s}(k_\ast) \simeq 2.1 \times 10^{-9}$. Moreover, the dimensionless parameter $\alpha$ can be in the range of $10^{-2.5}\leq \alpha\leq 10$, and it is set to $\alpha=0.005$ in our subsequent calculations \cite{Planck:2018inflation,Planck:2018inflation1}.

\begin{figure}[H]
	\centering
	\begin{minipage}[b]{1\textwidth}
		\centering
		\subfigure[\label{fig-phia1} ]{ \includegraphics[width=0.65\textwidth]%
			{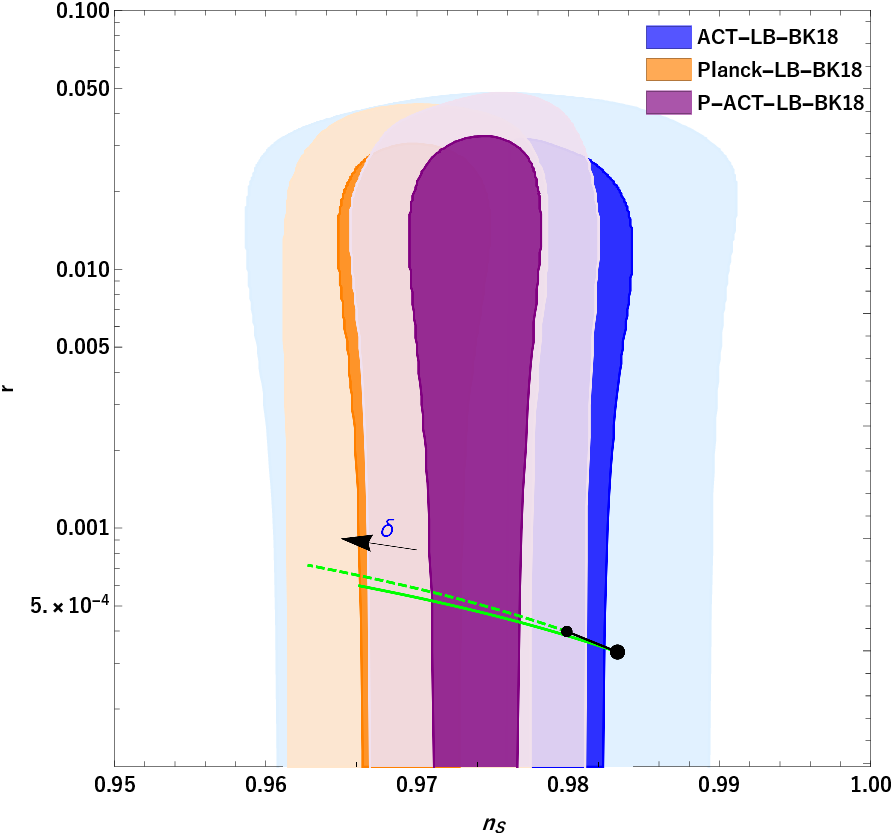}}
		\hspace*{0.02\textwidth}

	\end{minipage}
	\caption{The tensor-to-scalar ratio $r$ against the scalar spectral index $n_{\rm s}$ for the SUSY potential (\ref{eq:susy_pot}) with LQC inverse volume corrections. The dashed and solid green curves show the model predictions for $N=50$ and $N=60$ e-folds, respectively. For both cases, the parameters are fixed at $\alpha=0.005$ and $\sigma=2$, while $\delta$ is varied from $0$ to $3 \times 10^{-3}$.}
	\label{fig:susy_phi_evolution}
\end{figure}

In the classical case where LQC corrections are absent, $\delta=0$ in Eqs. (\ref{eq:ns_in_LQC}) and (\ref{eq:r_in_LQC}), the predictions for $r-n_{\rm s}$ diagram are shown by the black bar in  Fig.~\ref{fig:susy_phi_evolution}. It can be seen from this figure that for an $e$-fold range of $50\leq N\leq60$, the predictions lie outside the 68\% CL region of the P-ACT-LB-BK18 data. While the newest data brings the model closer to viability, it remains in tension with the most constrained 68\% CL region.

\begin{figure}[H]
		\centering
	\begin{minipage}[b]{1\textwidth}
		\centering
		\subfigure[\label{fig-phase_space_susy-N=50} ]{ \includegraphics[width=0.47\textwidth]%
			{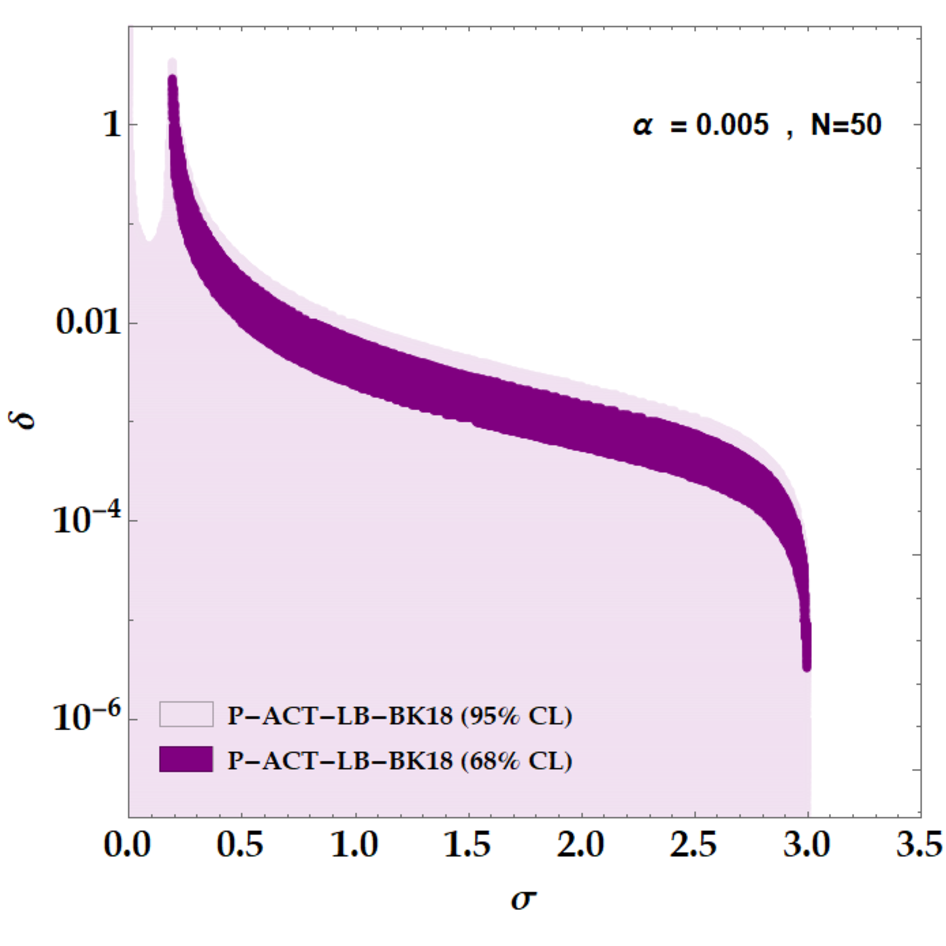}}
		\hspace*{0.02\textwidth}
		\subfigure[\label{fig-phase_space_susy-N=60} ]{ \includegraphics[width=0.47\textwidth]%
			{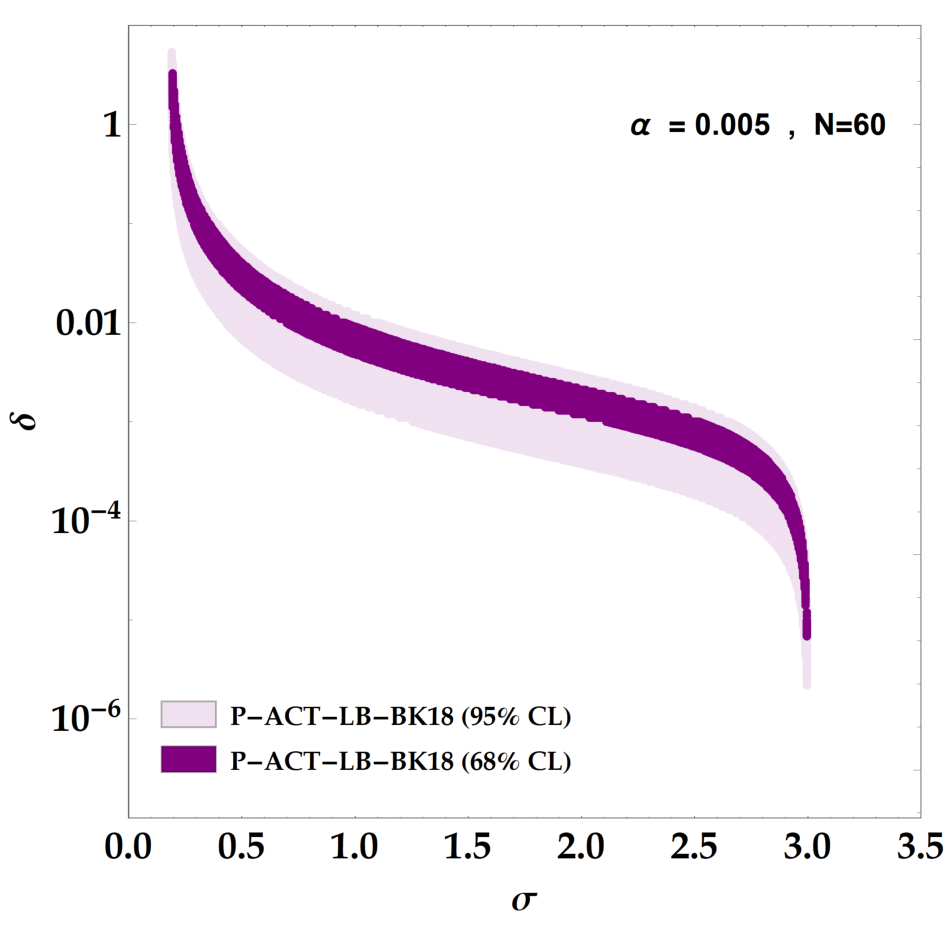}}
	\end{minipage}

	\caption{The allowed zones for the LQC inverse volume parameters $\sigma$ and $\delta$ in the phase space, for the SB SUSY potential (\ref{eq:susy_pot}), calculated for: (a) $N=50$ and (b) $N=60$. The plot is bounded by P-ACT-LB-BK18 data at 68\% CL (dark purple) and 95\% CL (light purple).}
		\label{fig:phase_space_susy}
	\end{figure}

The inclusion of inverse volume corrections significantly alters the predictions.
As shown in Fig. \ref{fig:susy_phi_evolution} the LQC corrections, primarily through the $c_{n_{\rm s}}\delta_{\rm pl}$ term in the expression for $n_{\rm s}$ in Eq. (\ref{eq:ns_in_LQC}), shift the predicted values horizontally to the left (see the green line).
By varying the LQC parameter $\delta$ for a fixed $\sigma$, the model predictions can be moved into the observationally favored 68\% and 95\% CL regions. The analysis of Fig. \ref{fig:susy_phi_evolution} reveals the impact of inverse volume corrections on the observational viability of the SB SUSY potential. The primary effect of the LQC corrections is to modify the scalar spectral index $n_{\rm s}$ while having a sub-dominant effect on the tensor-to-scalar ratio $r$. This is consistent with the theoretical framework, where the correction term in the equation for $n_{\rm s}$ (\ref{eq:ns_in_LQC}) induces a more significant shift than the corresponding term for $r$ (\ref{eq:r_in_LQC}). Consequently, the model predictions translate almost horizontally from right to left in the $r-n_{\rm s}$ plane as the LQC correction parameter $\delta$ increases for a given constant $\sigma$. In the limit of a vanishingly small correction $(\delta\leq10^{-7})$, the model predictions converge to the classical case, which is in tension with the observational data. As $\delta$ is increased, the predictions shift leftward that allows them to overlap with the P-ACT-LB-BK18 data contours. This behavior makes it possible to derive quantitative constraints on the LQC parameter space by identifying the range of $\delta$ that falls within the 95\% and 68\% confidence level regions for a given value of $\sigma$.
Table~\ref{tab:susy}
shows the permitted values for the LQC inverse volume parameter $\delta$ for different values of parameter $\sigma$ and inflationary $e$-folds number $N=50$ and $N=60$, in 68\% and 95\% CL of P-ACT-LB-BK18 data. Figure~\ref{fig:phase_space_susy} summarizes the viable parameter space in the  $\sigma-\delta$ plane. The allowed regions are delineated by the 68\% and 95\% CL constraints from the combined P-ACT-LB-BK18 dataset.
\begin{table}[h!]
\centering
\caption{Allowed ranges for the LQC inverse volume parameter $\delta$ with varying $\sigma$ for the SUSY potential (\ref{eq:susy_pot}), based on 68\% CL and 95\% CL constraints from P-ACT-LB-BK18 data for $e$-fold numbers $N=50$ and $N=60$.}
\label{tab:susy}
{\footnotesize
\begin{tabular}{c  | c @{\hspace{10pt}} c | c @{\hspace{10pt}} c}
\thickhline
& \multicolumn{2}{c|}{$N=50$} & \multicolumn{2}{c}{$N=60$} \\
\cline{2-5}
$\sigma$ & $\delta$ (95\% CL) & $\delta$ (68\% CL) & $\delta$ (95\% CL) & $\delta$ (68\% CL) \\
\thickhline
0.5 & $\delta \leq 4.2 \times 10^{-2}$ & $[9.6 \times 10^{-3}, 2.9 \times 10^{-2}]$ & $[6.4 \times 10^{-3}, 5.4 \times 10^{-2}]$ & $[2.0 \times 10^{-2}, 4.0 \times 10^{-2}]$ \\
1 & $\delta \leq 9.5 \times 10^{-3}$ & $[2.3 \times 10^{-3}, 6.6 \times 10^{-3}]$ & $[1.5 \times 10^{-3}, 1.2 \times 10^{-2}]$ & $[4.5 \times 10^{-3}, 9.3 \times 10^{-3}]$ \\
2 & $\delta \leq 2.2 \times 10^{-3}$ & $[5.5 \times 10^{-4}, 1.6 \times 10^{-3}]$ & $[3.6\times 10^{-4}, 2.9 \times 10^{-3}]$ & $[1.2 \times 10^{-3}, 2.2 \times 10^{-3}]$ \\
2.99 & $\delta \leq 2.6 \times 10^{-5}$ & $[6.4 \times 10^{-6}, 1.8 \times 10^{-5}]$ & $[3.6 \times 10^{-5}, 3.4 \times 10^{-4}]$ & $[1.2 \times 10^{-4}, 2.4 \times 10^{-4}]$ \\
\thickhline
\end{tabular}}
\end{table}

\section{Exponential potential in LQC}\label{sec:exp}

\begin{figure}[H]
	\centering
	\begin{minipage}[b]{1\textwidth}
		\centering
		\subfigure[\label{fig-exp-rns1} ]{ \includegraphics[width=0.47\textwidth]%
			{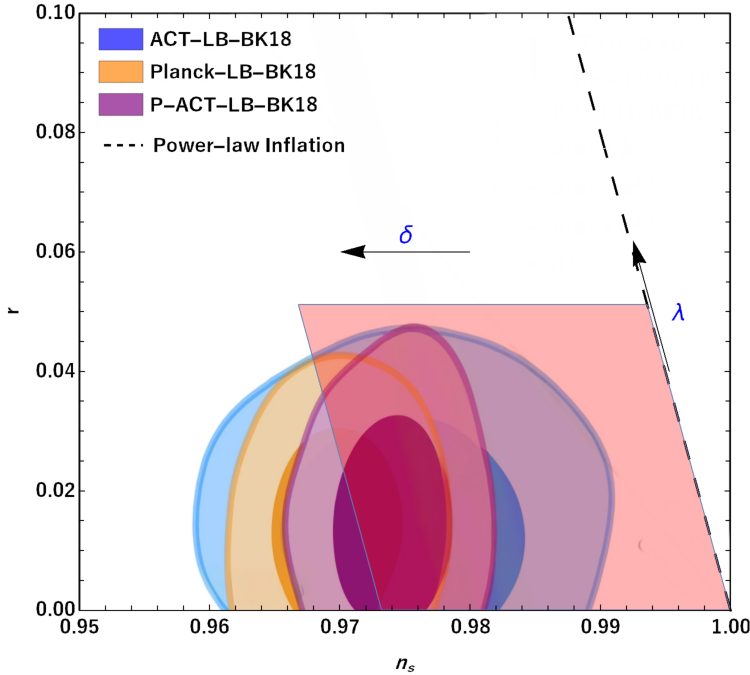}}
		\hspace*{0.02\textwidth}
		\subfigure[\label{fig-exp-rns2} ]{ \includegraphics[width=0.47\textwidth]%
			{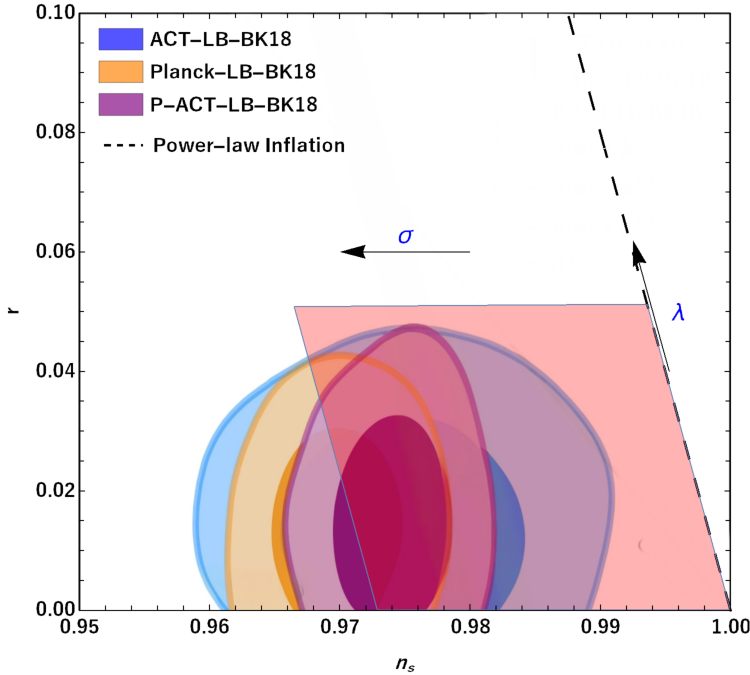}}
	\end{minipage}
	\caption{The $r-n_{\rm s}$ diagram for the exponential potential (\ref{eq:exp_potential}) in the LQC framework with inverse volume corrections. (a) The parameter $\delta$ is varied over the range $[0, 10^{-2}]$ for a fixed $\sigma=1$. (b) The parameter $\sigma$ is varied over the range $[0, 3)$ for a fixed $\delta = 10^{-3}$. In both panels, the black arrows indicate the direction of increase for the varying parameters. The dashed line curves show the prediction of standard power law inflation (i.e. without inverse volume corrections).}\label{fig:exp2_evolution}
\end{figure}

In this section, the behavior of the exponential potential within the context of LQC with inverse volume corrections is analyzed. The potential is given by
  \begin{align}\label{eq:exp_potential}
 	V(\varphi) = V_0 e^{-\lambda\left(\frac{\varphi}{M_{\rm p}}\right)},
 \end{align}
where $\lambda>0$. The constant $V_0$ is fixed by normalizing the scalar power spectrum to its observed value at the pivot scale. A key feature of this potential is that its corresponding slow roll parameters are constants and depend only on $\lambda$ as $\epsilon_{\rm V}=\lambda^2/2$ and $\eta=\lambda^2$. In the standard inflation (i.e., without quantum corrections), the exponential potential corresponds to the power-law inflation, where $a(t)\propto t^ n$ with $n> 1$, and is observationally ruled out \cite{Rezazadeh:2017,Amani:2018,Lucchin:1985,Yokoyama :1987,Burd:1988,Halliwell:1987}. As shown by the dashed line in Figs. \ref{fig-exp-rns1} and \ref{fig-exp-rns2}, its predictions in the $(r-n_{\rm s})$ plane lie far outside the 95\% CL region of the P-ACT-LB-BK18 dataset. Accordingly, this potential is not considered as a viable candidate in the classical framework.

Now, this potential is investigated within the LQC framework by considering the effects of inverse volume corrections on the inflationary perturbations. As established in Section
\ref{sec2}, the LQC corrections primarily modify the scalar spectral index $n_{\rm s}$ (\ref{eq:ns_in_LQC}), which can shift the predictions in the $r-n_{\rm s}$ plane horizontally. This provides a mechanism to move the model predictions into the observationally allowed regions. Using Eqs. (\ref{eq:ns_in_LQC}) and (\ref{eq:r_in_LQC}), we plot the behavior of the exponential potential (\ref{eq:exp_potential}) in the $r-n_{\rm s}$ plane for fixed values of $\sigma$ and $\delta$ in Figs. \ref{fig-exp-rns1} and \ref{fig-exp-rns2}, respectively. Figures show that for a given $\lambda$, increasing the LQC parameters $\sigma$ and $\delta$ shifts the predictions of the model to the left along the $n_{\rm s}$ axis. This allows us to constrain the LQC parameter space for which the exponential potential is observationally viable.

The analysis of the exponential potential (\ref{eq:exp_potential}) in the $r-n_{\rm s}$ plane reveals that the viability of the model is highly sensitive to the potential parameter $\lambda$. The results, summarized in Table \ref{tab:combined} shows i) for $\lambda > 0.076$, the model cannot be reconciled with the data for any values of the LQC parameters $\sigma$ and $\delta$; ii) for $0.064\leq\lambda \leq 0.076$ the model predictions can be shifted into the 95\% CL region, but they remain outside the more stringent 68\% CL contour; iii) for $\lambda < 0.064$, the potential becomes viable, as the LQC corrections can shift the $(r-n_{\rm s})$ predictions into the 95\% CL and 68\% CL regions.

\begin{figure}[H]
	\centering
	\begin{minipage}[b]{1\textwidth}
		\centering
		\subfigure[\label{phase-space-Lambda-01} ]{ \includegraphics[width=0.47\textwidth]%
			{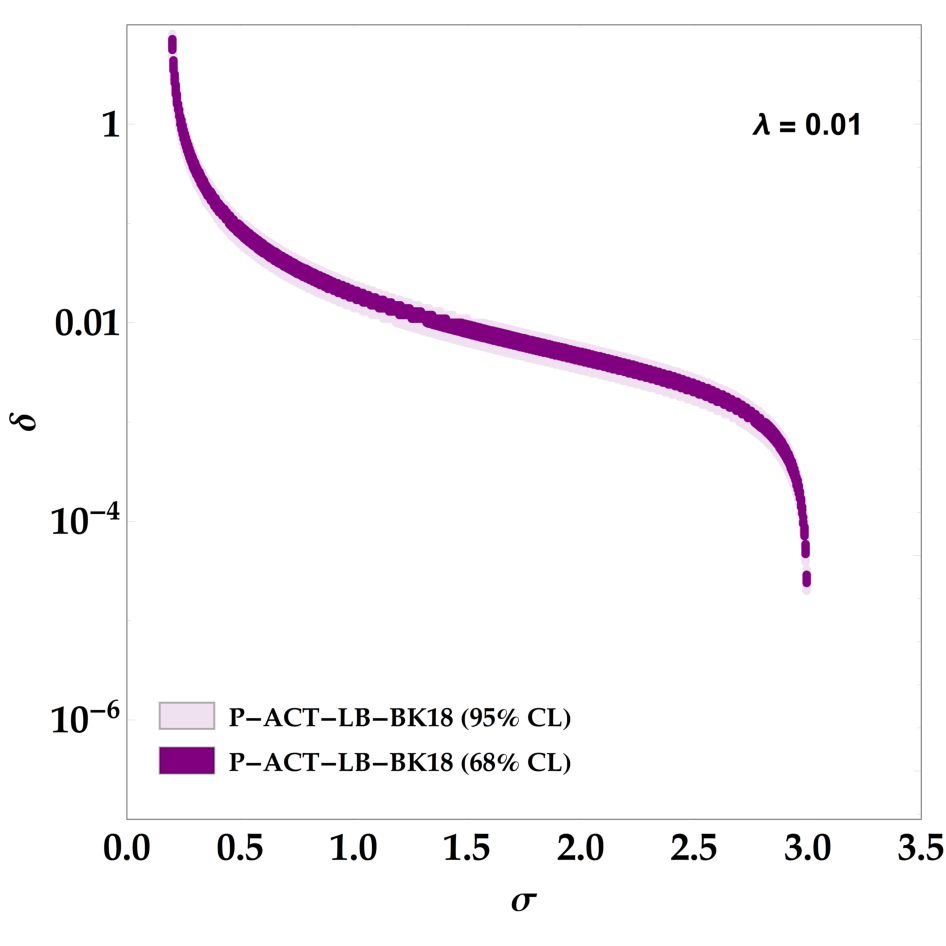}}
		\hspace*{0.02\textwidth}
		\subfigure[\label{phase-space-Lambda-04} ]{ \includegraphics[width=0.47\textwidth]%
			{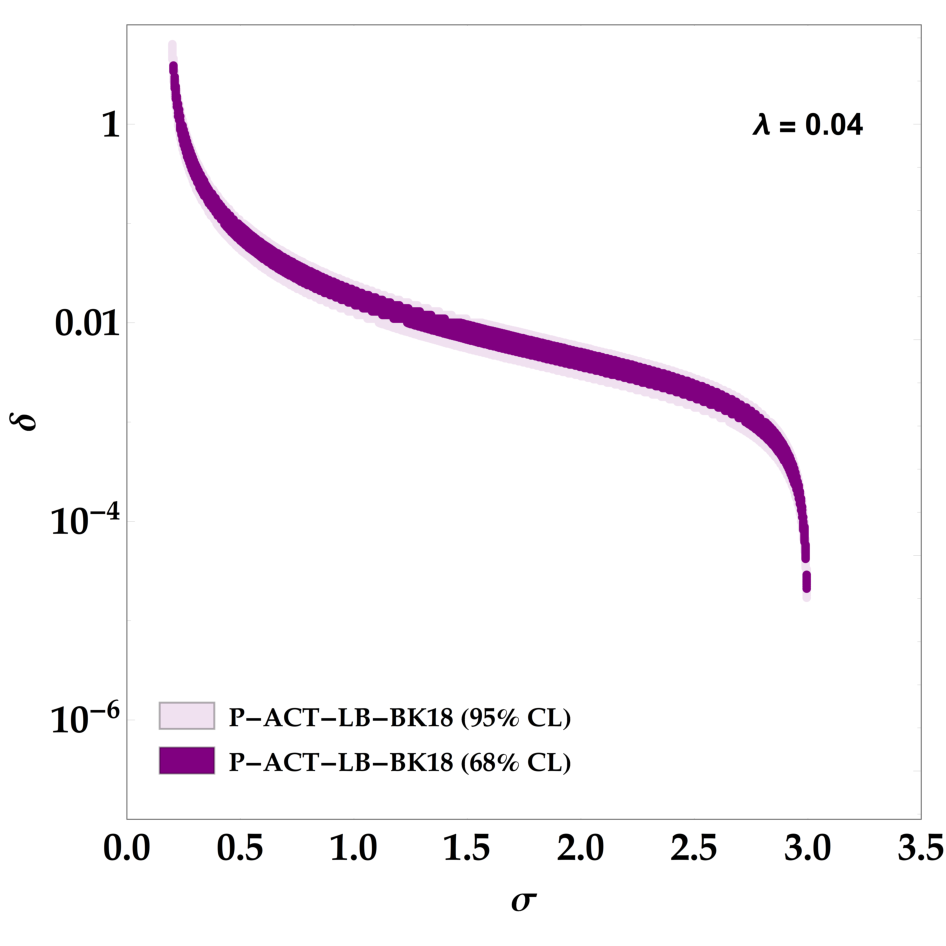}}
			\subfigure[\label{phase-space-Lambda-064} ]{ \includegraphics[width=0.47\textwidth]%
				{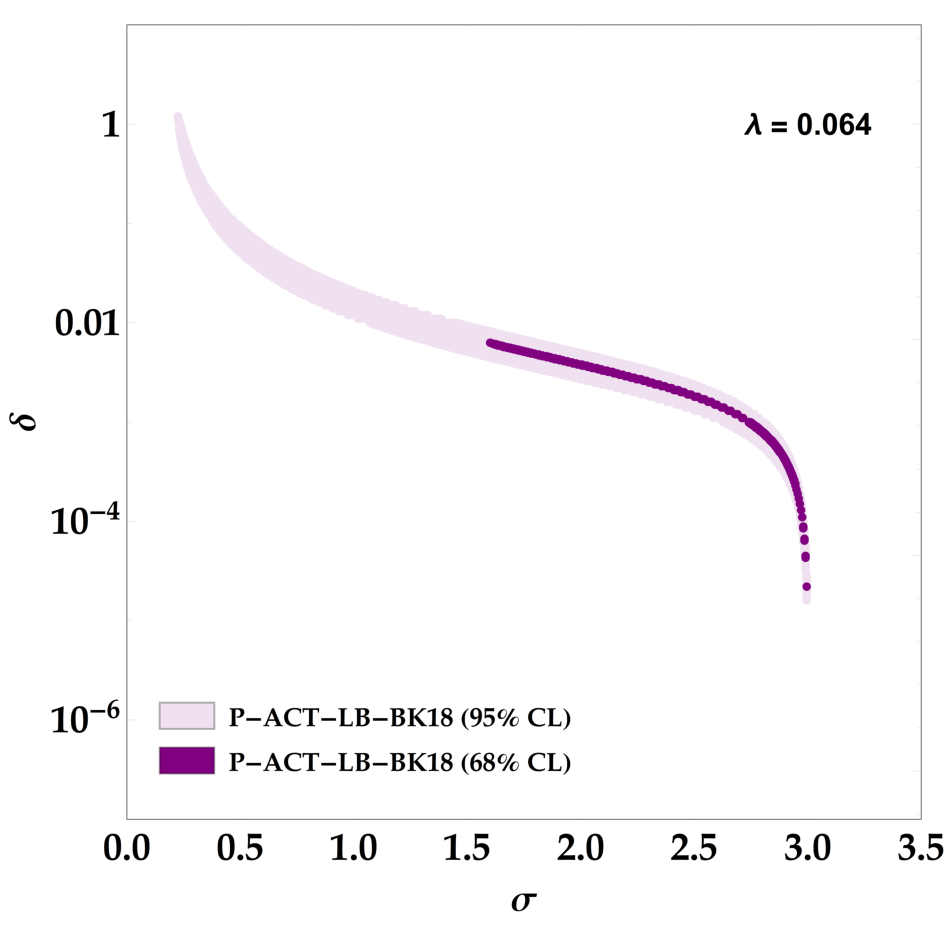}}
				\subfigure[\label{phase-space-Lambda-076} ]{ \includegraphics[width=0.47\textwidth]%
					{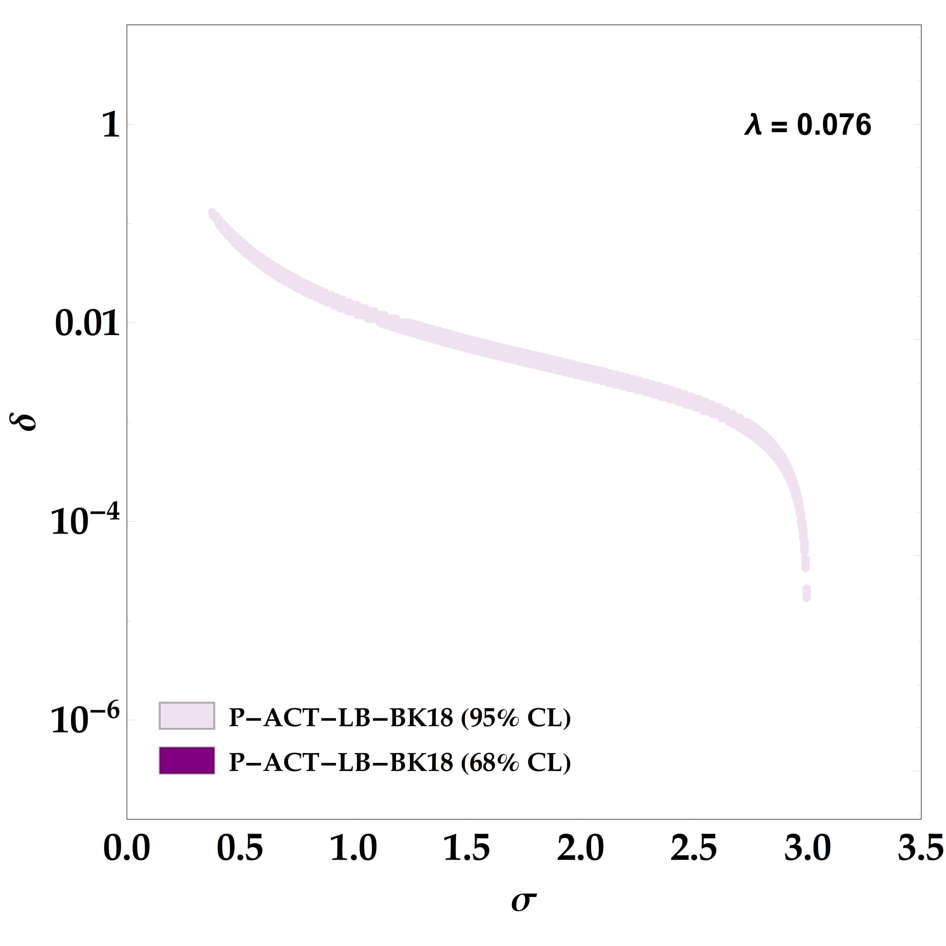}}
	\end{minipage}
	\caption{The allowed regions for the LQC parameters $\sigma$ and $\delta$ in the phase space for the exponential potential (\ref{eq:exp_potential}), constrained by P-ACT-LB-BK18 data at 68\% (dark purple) and 95\% (light purple) confidence levels. The panels show the results for different values of $\lambda$: (a) $\lambda=0.01$, (b) $\lambda=0.04$, (c) $\lambda=0.064$, and (d) $\lambda=0.076$.}\label{fig:exp_phase-space}
\end{figure}

Figures \ref{fig-exp-rns1} and \ref{fig-exp-rns2} illustrate the behavior of the model for fixed values of $\sigma$ and $\delta$, respectively.
For a given $\lambda$, increasing the LQC parameters $\delta$ and $\sigma$ shifts the predictions to the left along the $n_{\rm s}$ axis. This allows us to constrain the LQC parameter space for which the exponential potential is observationally viable. Figure \ref{fig:exp_phase-space} presents a comprehensive phase-space diagrams for the allowed regions in the $(\sigma -\delta)$ plane, for different $\lambda$, constrained by the P-ACT-LB-BK18 dataset at both 68\% and 95\% confidence levels.


\begin{table}[H]
\centering
\caption{Allowed ranges for the parameter $\delta$ for different values of $\sigma$ and $\lambda$, based on 68\% CL and 95\% CL constraints from P-ACT-LB-BK18 data \cite{ACT:DR6models} for the exponential potential (\ref{eq:exp_potential}).}\label{tab:combined}
{\footnotesize
\begin{tabular}{c | c @{\hspace{20pt}} c  @{\hspace{10pt}}|  @{\hspace{10pt}}c @{\hspace{20pt}} c}
\thickhline
& \multicolumn{2}{c@{\hspace{10pt}}|@{\hspace{10pt}}}{$\sigma = 1$} & \multicolumn{2}{c}{$\sigma = 2$} \\ \cline{2-5}
$\lambda$ & $\delta\times10^{-2}$ (95\% CL) & $\delta\times10^{-2}$ (68\% CL) & $\delta\times10^{-3}$ (95\% CL) & $\delta\times10^{-3}$ (68\% CL) \\ \thickhline
$0.01 $ & $[1.3, 2.5]$ & $[1.6, 2.2]$ & $[3.2, 6.0]$ & $[3.8, 5.2]$ \\
$0.04 $ & $[1.2, 2.5]$ & $[1.4, 2.2]$ & $[2.7, 5.9]$ & $[3.4, 5.2]$ \\
$0.064$ & $[1.1, 2.1]$ & $-$ & $[2.5, 5.2]$ & $[3.5,3.9]$ \\
$0.076$ & $[1.2, 1.5]$ & $-$ & $[3.0, 3.6]$ & $-$ \\
\thickhline
\end{tabular}}
\end{table}
\section{Conclusions}\label{sec:conclusion}
Within the framework of LQC, we have studied the SB SUSY and exponential inflationary potentials by incorporating the effects of inverse volume corrections. We demonstrated that while the scalar spectral index $n_{\rm s}$ and tensor-to-scalar ratio $r$ predicted by the classical versions of these models are in tension with the latest observational data from ACT, the inclusion of quantum gravitational effects can significantly improve their viability. Confronting the LQC-corrected predictions with the P-ACT-LB-BK18 dataset, yields the following outcomes:

\begin{itemize}
    \item In the absence of quantum corrections, both the SB SUSY and exponential potentials yield predictions for the $(r- n_s)$ plane that fall outside the 68\% confidence level region of recent data which challenge their observational viability.

    \item The inclusion of LQC inverse volume corrections provides a physically motivated mechanism to reconcile these potentials with observations. The dominant effect of these corrections is a negative shift in the scalar spectral index, $n_s$, which horizontally displaces the model predictions into the observationally favored regions of the parameter space.

    \item For the SB SUSY potential, the allowed ranges for the LQC parameters, $\sigma$ and $\delta$, were derived. As detailed in Table \ref{tab:susy}, a broad region of the parameter space was found to be consistent with the 68\% and 95\% confidence level contours of the data.

     \item The exponential potential, which is strongly disfavored in the classical context, can be revived observationally by LQC corrections. However, this is only possible for sufficiently small values of the potential parameter, specifically for $\lambda\leq 0.076$. For larger values, the model cannot be reconciled with the data for any choice of LQC parameters. The specific constraints on the LQC parameters for these viable scenarios are presented in Table \ref{tab:combined}.

\end{itemize}
In summary, the SUSY and exponential potentials are compatible with current observational data for specific ranges of the LQC parameters. The inverse volume corrections offer a supplementary layer of phenomenological richness, which demonstrates the significance of combining high-precision observational data with theoretical bounds from quantum gravity to improve and constrain the parameter space of inflationary scenarios.

\end{document}